\documentclass[12pt,a4j]{article}
\usepackage[dvips]{graphicx}
\usepackage{enumerate}
\usepackage{amsmath}
\usepackage{amssymb}
\usepackage{amsfonts}

\begin{document}
\baselineskip=6mm
\centerline{\bf  Parametric  solutions of the generalized short pulse equations  }\par
\bigskip
\centerline{Yoshimasa Matsuno\footnote{{\it E-mail address}: matsuno@yamaguchi-u.ac.jp}}\par
\centerline{\it Division of Applied Mathematical Science,}\par
\centerline{\it  Graduate School of Sciences and Technology for Innovation,} \par
\centerline{\it Yamaguchi University, Ube, Yamaguchi 755-8611, Japan} \par
\bigskip
\bigskip
\noindent{\bf Abstract}\par
\noindent We consider three novel PDEs associated with the integrable generalizations of the short pulse equation classified recently
by Hone {\it et al} (2018 {\it Lett. Math. Phys.} {\bf 108} 927-947).
In particular, we obtain a variety of exact solutions  by means of a direct method analogous to that used for solving the short pulse equation. 
The main results reported here are the  parametric representations  of the multisoliton solutions.
These solutions include cusp solitons, unbounded  solutions with finite slope  and breathers. 
 In addition, the cusped periodic wave solutions are constructed from the cusp solitons 
 by means of a simple procedure.
 As for non-periodic solutions, smooth breather solutions are of particular interest from the perspective of applications to real physical phenomena.
 The cycloid reduced from the periodic traveling wave  with cusps is also  worth remarking in connection with  Gerstner's trochoidal solution in deep gravity waves.
A number of works are left for future study, some of which will be  addressed in concluding remarks.
\par
\bigskip

\newpage
\leftline{\bf  1. Introduction} \par
\medskip
\noindent The short pulse (SP) equation is a completely integrable partial differential equation (PDE) in the
sence that it admits a Lax pair and an infinite number of conservation laws [1].
It can be written in an appropriate dimensionless form as
$$u_{xt}=u+{1\over 6 }(u^3)_{xx}, \eqno(1.1)$$
where $u=u(x,t)$ represents  a scalar function of $x$ and $t$, and subscripts $x$ and $t$
appended to $u$ denote  partial differentiations.
The SP equation has been found for the first time
in an attempt to construct integrable differential equations associated with pseudospherical surfaces [2, 3].
In the physical context, it has been derived as an asymptotic model describing the propagation of ultra-short  pulses in isotropic optical fibers [4].
A large number of works have been devoted to the study of the SP equation. See, for example, a review article [5] as for the soliton and periodic solutions
and their properties. \par
Quite recently, Hone  {\it et al}. [6] classified  the integrable PDEs of second order  with quadratic and cubic nonlinear terms.
They found seven integrable PDEs including the SP equation.  The remaining six equations read
$$u_{xt}=u+(u^2)_{xx}, \eqno(1.2)$$
$$u_{xt}=u+2uu_{xx}+u_x^2, \eqno(1.3)$$
$$u_{xt}=u+u^2u_{xx}+uu_x^2, \eqno(1.4)$$
$$u_{xt}=u+4uu_{xx}+u_x^2, \eqno(1.5)$$
$$u_{xt}=u+(u^2-4u^2u_x)_x, \eqno(1.6)$$
$$u_{xt}=u+\alpha(2uu_{xx}+u_x^2)+\beta(u^2u_{xx}+uu_x^2), \quad \alpha\beta\not=0. \eqno(1.7)$$
Equations (1.2), (1.3) and (1.4) are already  known  and  sometimes called the Vakhnenko (or reduced Ostrovsky), Hunter-Suxton and modified  SP equations, respectively [7-10].
Recall that equations (1.2) and (1.3) stem from the short-wave limit of the   Degasperis-Procesi and Camassa-Holm  equations, respectively,
and their parametric solutions have been obtained.  See, for
example Matsuno [11].  While equations (1.3) and (1.5) have been derived  in the process of studying the short-wave dynamics of surface gravity waves [12], 
  the analysis of the latter equation has not been done yet. On the other hand,
 equations (1.6) and  (1.7) seem to be new, as far as we know. 
 \par
The integrability of the above PDEs was established by constructing the Lax pairs and recursion operators.
In addition, by means of the reciprocal transformation, equations (1.5) and (1.6) were shown to be transformed to the Tzitzeica equation
whereas equation (1.7) was found to be related to the sine-Gordon (sG) equation [6]. See also an analogous work demonstrating that equation (1.7) is
transformed to the sG equation by means of the reciprocal transformation combined with a sequence of dependent variable transformations [13].
These reciprocal links between equations (1.5)-(1.7) and the integrable Tzitzeica  and sG equations  suggest the existence of exact solutions of the
former equations. Nevertheless, their construction still remains  as an open problem. 
The purpose of this paper is to present the parametric representations  for exact solutions of equations (1.5)-(1.7) and investigate their properties.
The exact method of solution employed here is similar to that used in constructing soliton solutions of the SP and modified SP equation and does not need the
knowledge of the inverse scattering transform (IST)  method [10, 14]. \par
The present paper is organized as follows. In section 2, we present the parametric representation of the solutions of equation (1.5), and show that the solutions take the form of cusp solitons.
The cusped periodic wave solutions are obtained simply from the non-periodic cusp solitons by setting the soliton parameters to be pure imaginary.
The tau-function for the cusp soliton solutions is   closely related to that derived from a scaling limit  of the tau-function
associated with the soliton solutions of the  Novikov equation [15].  This observation makes it possible to obtain
the parametric  solutions in terms of a single tau-function. In section 3, the similar procedure to that used for equation (1.5) is applied to
equation (1.6) to obtain the parametric solutions.  The soliton solutions exhibit peculiar features. In particular, the solutions are smooth, but they always  diverge irrespective of the values of 
the soliton parameters. 
The degree of divergence is, however  found to be not so bad  that the $x$ derivative of the solutions remains finite, and indeed takes the form of either kink or soliton.
The smooth and  bounded solutions have not been obtained yet as long as non-periodic solutions are concerned. In section 4, we first recast  equation (1.7) to a PDE which is quite similar to the modified SP equation (1.4) 
by means of the linear transformation. Then, the equation  is transformed to the modified sG equation through a sequence of  nonlinear dependent variable 
transformations. The tau-functions for the kink solutions of the modified sG equation as well as the bilinear equations that they satisfy are used to obtain the parametric solutions.
The solutions thus constructed are classified into two groups. One belongs to singular cusp  solitons and their periodic analogs, and the other one is characterized by breathers.  The former
solutions include the cycloid which is reduced from the  cusped periodic wave  by means of a limiting procedure. As for the latter solutions, 
we found  both smooth and singular breather solutions depending on the amplitude parameter of the breather.  The properties of solutions  are explored in some detail. 
In particular, the condition for the existence of smooth breathers  is derived. Section 5 is devoted to concluding remarks where we 
comment on some open problems associated with three PDEs under consideration.  
\par
\bigskip
\leftline{\bf 2. Parametric solutions of equation (1.5)}\par
\medskip
\noindent First, we rewrite equation (1.5) in the form
$$u_{xt}={3\over 2}\,u-uu_{xx}-{1\over 4}\,u_x^2, \eqno(2.1)$$
by rescaling the variables according to $t\rightarrow (3/2)t, u\rightarrow -u/6$. Then, equation (2.1) can be put into the
local conservation law
$$p_t+(pu)_x=0, \quad p=m^{2\over 3}, \quad m=1-u_{xx}. \eqno(2.2)$$
We construct solutions of equation (2.1) under the boundary condition $u \rightarrow 0,\ |x| \rightarrow \infty$. \par
The goal in this section is to establish the following theorem. \par
\medskip
\noindent {\bf Theorem 2.1.} {\it Equation (2.1) admits the parametric representation for the multisoliton solutions 
$$u=-2({\rm ln}\,g)_{\tau\tau}, \eqno(2.3a)$$
$$x=y-2({\rm ln}\,g)_\tau+y_0, \eqno(2.3b)$$
where $g=g(y, \tau)$ is the tau-function given by
$$g=f^2-2D_\tau D_yf\cdot f, \eqno(2.4)$$
with $f=f(y, \tau)$ being the fundamental tau-function expressed by a finite sum
$$ f=\sum_{\mu=0,1}{\rm exp}\left[\sum_{i=1}^N\mu_i\left(\xi_i+\pi{\rm i}\right)
+\sum_{1\le i<j\le N}\mu_i\mu_j\gamma_{ij}\right], \eqno(2.5a)$$
$$\xi_i=k_iy+{3\over 2k_i}\,\tau+\xi_{i0}, \quad (i=1, 2, ..., N),\eqno(2.5b)$$
$${\rm e}^{\gamma_{ij}}={(k_i-k_j)^2(k_i^2-k_ik_j+k_j^2)\over(k_i+k_j)^2(k_i^2+k_ik_j+k_j^2)}, \quad (i, j=1, 2, ..., N; i\not=j). \eqno(2.5c)$$
Here, $k_i$ and $\xi_{i0}$  are arbitrary complex  parameters, 
 and $N$ is an arbitrary positive integer. The notation $\sum_{\mu=0,1}$
implies the summation over all possible combinations of $\mu_1=0, 1, \mu_2=0, 1, ..., 
\mu_N=0, 1$.
The bilinear operators $D_\tau$ and $D_y$ in (2.4) are defined by
$$D_\tau^mD_y^nf\cdot g=\left({\partial\over\partial \tau}-{\partial\over\partial \tau^\prime}\right)^m
\left({\partial\over\partial y}-{\partial\over\partial y^\prime}\right)^n
f(y, \tau)g( y^\prime, \tau^\prime)\Big|_{\tau^\prime=\tau,\, y^\prime=y},\quad (m, n = 0, 1, 2, ...).  $$
}
\par
\bigskip
Recall that the tau-function $f$ from (2.5) is the same  functional form as that of the $N$-soliton solution of the Sawada-Kotera equation [16].
A factor $\pi{\rm i}$ in the exponential functions in $(2.5a)$ has been introduced to assure the boundedness of the solutions. 
The above expression for $u$ will be shown to represent the non-periodic $N$-cusp soliton solution
for the real parameters $k_j$ and $\xi_{j0}$. The former parameters are related to the amplitudes of solitons whereas the latter ones represent the phases of solitons.
The cusped periodic wave solutions can be constructed by replacing these real parameters with the pure imaginary ones, i.e., $k_j\rightarrow {\rm i}k_j, \xi_{j0}\rightarrow {\rm i}\xi_{j0}\ (j=1, 2, ..., N)$.
Theorem 2.1 is proved by a sequence of steps. We start from the reciprocal transformation of equation (2.2) under the boundary condition $u\rightarrow 0, |x|\rightarrow \infty$. \par
\bigskip
\leftline{\bf 2.1.  Reciprocal transformation}\par
\medskip
\noindent Equation (2.2) enables us to introduce the reciprocal transformation $(x, t) \rightarrow (y, \tau)$ by
$$dy=pdx-pudt, \quad d\tau=dt. \eqno(2.6)$$
This leads to the system of linear PDEs for the variable $x=x(y, \tau)$
$$x_y={1\over p}, \quad x_\tau=u. \eqno(2.7)$$
It follows from the compatibility condition of the above system that
$$\left({1\over p}\right)_\tau=u_y. \eqno(2.8)$$
In view of  (2.8), one can express $m$ from (2.2) in terms of $p$ as
$$m=1-p^2u_{yy}-pp_yu_y=1-p^2\left({1\over p}\right)_{\tau y}-pp_y\left({1\over p}\right)_\tau= 1+p({\rm ln}\,p)_{\tau y}. \eqno(2.9)$$
Since $m=p^{3/2}$ by the definition,  (2.9) yields a single nonlinear PDE for $p$
$$({\rm ln}\,p)_{\tau y}=p^{1\over 2} - p^{-1}. \eqno(2.10)$$
This equation is a form of the Tzitzeica equation [6] which is also  referred to  as the Bullough-Dodd equation [17, 18]. \par
\bigskip
\leftline{\bf 2.2  Parametric  representation of  solutions} \par
\medskip
\noindent We introduce the new variables $U$ and $W$ by
$$U=-{p_{yy}\over 2p}+{p_y^2\over 4p^2}, \quad W=p^{1\over 2}. \eqno(2.11)$$
Using Eq. (2.10), one can show that $U_\tau=-3p_y/(4p^{1/2})$ which,  combined with the relation $W_y=p_y/(2p^{1/2})$, gives
$$U_\tau+{3\over 2}\,W_y=0. \eqno(2.12)$$
On the other hand, it follows by a direct computation with  use of (2.11) that
$$W_{yy}+UW=0. \eqno(2.13)$$
By eliminating the variable $W$ from (2.12) and (2.13), one arrives at the single nonlinear PDE for $U$
$$UU_{\tau yy}-U_yU_{\tau y}+U^2U_\tau=0. \eqno(2.14)$$
\par
As shown below, equation (2.14) can be bilinearized and admits multisoliton solutions. 
Indeed, introducing the dependent variable transformation with the tau-function $f$
$$U=6({\rm ln}\,f)_{yy}, \eqno(2.15)$$
   equation (2.14) can be integrated once with respect to $y$, giving
$${U_{\tau y}\over U}+6({\rm ln}\,f)_{\tau y}=c, \eqno(2.16)$$
where $c$ is an integration constant. 
The relation $U_{\tau y}/U=3W/2$ which follows from (2.12) and (2.13) yields the limiting value
$${\rm lim}_{|y|\rightarrow\infty}{U_{\tau y}\over U} = {\rm lim}_{|y|\rightarrow\infty} {3\over 2}\,W={\rm lim}_{|y|\rightarrow\infty}\,{3\over 2}\,p^{1\over 2}={3\over 2},  \eqno(2.17)$$
where we have used  (2.11) for $W$ and (2.2) for $p$ as well as 
 the boundary condition $u\rightarrow 0, |x|\rightarrow \infty$ in the expression of $p$.
 We also assume that ${\rm lim}_{|y|\rightarrow\infty}({\rm ln}\,f)_{\tau y}=0$. Taking the limit $|y|\rightarrow\infty$ in (2.16) with these limiting values in mind, one
 finds that  $c=3/2$.   Equation (2.16) then becomes
$$U_{\tau y}+6({\rm ln}\,f)_{\tau y}U={3\over 2}\,U. \eqno(2.18)$$
Substituting the variable $U$ from (2.15) into equation (2.18), we obtain the bilinear equation for $f$
$$D_\tau D_y^3f\cdot f-{3\over 2}\,D_y^2f\cdot f=0. \eqno(2.19)$$
\par
To clarify the structure of equation (2.19), we introduce the variable $q=q(y, \tau)$ by $q=3({\rm ln}\,f)_{yy}$. Then, equation (2.19) is transformed to a nonlocal
nonlinear PDE for $q$. It reads
$${3\over 2}\,q_y-3qq_\tau+3q_y\int_y^\infty\,q_\tau\,dy-q_{\tau yy}=0. \eqno(2.20)$$
Recall that equation (2.20) can be derived from the short-wave limit of a model equation for shallow-water waves [19]
$$q_\tau+{3\over 4}\,q_y-{3\over 2}\,qq_\tau+{3\over 2}\,q_y\int_y^\infty\,q_\tau\,dy-{1\over 2}\,q_{\tau yy}=0. \eqno(2.21)$$
Actually, equation (2.20) arises from equation (2.21) by rescaling the variables according to $y\rightarrow \epsilon y, \tau\rightarrow \tau/\epsilon, q\rightarrow q/\epsilon^2$
and taking the limit $\epsilon \rightarrow 0$.  The explicit form of the tau-function (2.3) can be derived by taking the same scaling limit from the tau-function of the
$N$-soliton solution of the shallow-water wave equation. See the expression (2.20) of [15].
If we substitute (2.15) into (2.12) and integrate it once with respect to $y$ 
under the boundary condition ${\rm lim}_{|y|\rightarrow\infty}W=1$, we obtain
$$W=1-4({\rm ln}\,f)_{\tau y}. \eqno(2.22)$$
In view of the definition of $g$ from (2.4), one can express $W$ in terms of the tau-functions $f$ and $g$ as
$$W={g\over f^2}. \eqno(2.23)$$
\par
The final step for the proof of Theorem 2.1 is provided by the following proposition. \par
\bigskip
\noindent {\bf Proposition 2.1.} {\it The tau-function $g$ from (2.4) satisfies the bilinear identity
$$g^2-D_\tau D_yg\cdot g=f^4. \eqno(2.24)$$}
\bigskip 
{\bf Proof.}  First, define the tau-functions $f_1, f_2, g_1$ and $g_2$ by the relations
$$f_1=f(\xi-\phi), \quad f_2=f(\xi+\phi), \quad g_1=\hat g(\xi-\phi), \quad g_2=\hat g(\xi+\phi), $$
where $\hat g=f_1f_2+\kappa D_yf_1\cdot f_2\ (\kappa=1/\sqrt{2})$ and $\xi=(\xi_1, \xi_2, ..., \xi_N)$ and $\phi=(\phi_1, \phi_2, ..., \phi_N)$ are $N$-component row vectors
with $\phi_i$ being given by
$${\rm e}^{-\phi_i}=\sqrt{{\left(1-{\kappa k_i\over2}\right)(1-\kappa k_i)\over \left(1+{\kappa k_i\over 2}\right)(1+\kappa k_i)}}, \quad (i=1, 2, ..., N). \eqno(2.25)$$
It then follows from the relation $(3.6a)$ of [15], that
$$\left(D_y+{2\over \kappa}\right)g_1\cdot g_2={2\over\kappa}f^4. \eqno (2.26)$$
Rescaling the variables as $y\rightarrow \epsilon y, \tau\rightarrow \tau/\epsilon, k_i\rightarrow k_i/\epsilon$ and taking the limit $\epsilon\rightarrow 0$, we find 
that at the leading-order, $\phi_i=3\epsilon/(\kappa k_i)+O(\epsilon^2)$.
The leading-order asymptotics of the
tau-functions $f_1, f_2, g_1$ and $g_2$  can then be expressed as
$$f_1=f-{2\epsilon\over\kappa}\,f_\tau+O(\epsilon^2), \quad f_2=f+{2\epsilon\over\kappa}\,f_\tau+O(\epsilon^2), \eqno(2.27)$$
$$g_1=g-{2\epsilon\over\kappa}\,g_\tau+O(\epsilon^2), \quad g_2=g+{2\epsilon\over\kappa}\,g_\tau+O(\epsilon^2). \eqno(2.28)$$
The relation (2.24) follows by substituting these asymptotics into (2.26) and taking the limit $\epsilon\rightarrow 0$.  \hspace{\fill}$\square$
\par
Finally, it follows from (2.7), (2.11), (2.23) and (2.24) that
$$x_y={1\over W^2}={f^4\over g^2}=1-2({\rm ln}\,g)_{\tau y}. \eqno(2.29)$$
Integrating the above equation under the boundary condition $x_\tau=u \rightarrow 0,  |y|\rightarrow \infty$, one arrives at the expression $(2.3b)$ for $x$
 with $y_0$ being an arbitrary constant. 
The expression $(2.3a)$ for $u$ is obtained simply from $(2.3b)$ and the second equation in (2.7).  This completes the proof of Theorem 2.1. \par
\bigskip
\leftline{\bf 2.3. Soliton and periodic solutions} \par
\medskip
\leftline{\bf 2.3.1. Cusp soliton} \par
\noindent Here, we explore the properties of soliton solutions for both the non-periodic and periodic cases. In particular, we show that soliton solutions take the form of cusp solitons.
The tau-functions for the   one-soliton solution are given by (2.4) and (2.5) with $N=1$ and $k_1, \xi_1\in \mathbb{R}$. They read
$$f=1-{\rm e}^\xi, \quad g=1+4{\rm e}^\xi+{\rm e}^{2\xi}, \quad \xi=ky+{3\over 2k}\,\tau+\xi_0, \eqno(2.30)$$
where we have put $\xi=\xi_1, k=k_1$ and $\xi_0=\xi_{10}$ for simplicity. The parametric representation of the solution follows from (2.3) and (2.30).
It can be written in the form
$$u=-{9\over 2k^2}{2\,\cosh\,\xi+1\over (\cosh\,\xi+2)^2}, \eqno(2.31a)$$
$$X\equiv x+{3\over 2k^2}\,t+x_0={\xi\over k}-{3\over k}{\sinh\,\xi\over \cosh\,\xi+2}, \eqno(2.31b)$$
where
 the traveling-wave coordinate $X$ has been introduced for convenience in which $x_0$ is an arbitrary real constant.
 A typical profile of $u$ is depicted on the left panel of figure 1 as a function of  $X$.
 It represents a cusp soliton with the amplitude $3/(2k^2)$ and the velocity $3/(2k^2)$.
  To see the structure of the singularity in more detail, we compute the $X$ derivative of $u$ from (2.31) and obtain
$$u_X={9\over k}{1\over \tanh\,{\xi\over 2}}{1\over \cosh\,\xi+2}. \eqno(2.32)$$
Near the trough $X=0\ (\xi=0)$, one estimates that $u_X\sim 6/(k\xi), X\sim \xi^5/(180k)$, and hence 
${\rm lim}_{X\rightarrow\pm 0}u_X={\rm lim}_{X\rightarrow\pm 0}(218/5k^6)^{1/5}/X^{1/5}=\pm\infty$, showing the appearance of the cusp.
 \par
 \medskip
\leftline{\bf 2.3.2.  Cusped periodic wave} \par
\noindent The cusped periodic wave solution is obtained formally from (2.31) if one replaces the parameters as $k\rightarrow {\rm i}k,  \xi\rightarrow {\rm i}\xi$.
It reads in the form
$$u={9\over 2k^2}{2\,\cos\,\xi+1\over (\cos\,\xi+2)^2}, \eqno(2.33a)$$
$$X\equiv x-{3\over 2k^2}\,t+x_0={\xi\over k}-{3\over k}{\sin\,\xi\over \cos\,\xi+2}. \eqno(2.33b)$$
The expression corresponding to (2.32) becomes
$$u_X=-{9\over k}{1\over \tan\,{\xi\over 2}}{1\over \cos\,\xi+2}. \eqno(2.34)$$
It represents the cusped periodic wave with the period $2\pi/k$. The cusp singularity appears
at the crest of the periodic wave.
The middle panel of figure 1 shows the profile of a cusped periodic wave with the period $2\pi\sqrt{2/3}$. \par
It is  important to observe that (2.33) belongs to a class of singular traveling waves. Actually, the traveling wave ansatz  $u=u(X),\ X=x-ct-x_0\ (c, x_0\in \mathbb{R})$
reduces Eq. (2.1) to an ODE $-cu_{XX}=3u/2-uu_{XX}-u_X^2/4$. Integrating this equation once with respect to $X$, we obtain
$$\left({du\over dX}\right)^2=2u-{4d\over |u-c|^{1/2}}+4c, \eqno(2.35)$$
where $d\in \mathbb{R}$ is an integration constant. One can confirm that (2.33) indeed satisfies Eq. (2.35) with $d=-c^{3/2}$ and $c=3/(2k^2)$. \par
 \medskip
\leftline{\bf 2.3.3.  Peaked periodic wave} \par
\noindent
Equation (2.35) exhibits smooth periodic solutions if $c>0$ and $0<d<c^{3/2}$. In particular, in the limit of $d\rightarrow c^{3/2}$, the
periodic wave reduces to an infinitesimal wave of the form $u\sim a\,\cos\sqrt{3/2c}X, \ |a|<<1,\ a\in\mathbb{R}$. In the limit  $d\rightarrow 0$,
on the other hand, it reduces to the peaked periodic wave with  parabolic profile
$$u={1\over 2}(X^2-4c), \quad X\in  [-\sqrt{6c}, \sqrt{6c}], \eqno(2.36)$$
periodically continued beyond the interval $ [-\sqrt{6c}, \sqrt{6c}]$. 
It should be remarked that the solution (2.36) can be derived directly by integrating Eq. (2.35) with $d=0$.
See the right panel of figure 1. 
The phase plane analysis of Eq. (2.35) exhibits a variety of periodic solutions. The detail
will be reported elsewhere. \par
\begin{center}
\includegraphics[width=16cm]{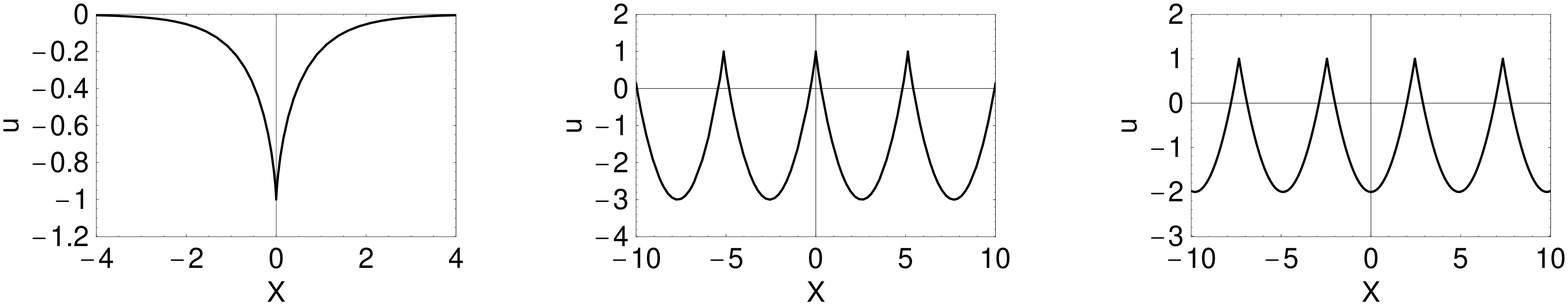}
\end{center}
\noindent {\bf Figure 1}\ The profiles of the cusp soliton, cusped periodic wave and peaked periodic wave solutions   with the parameter $k=\sqrt{3/2}$.
Left: Cusp soliton. Middle: Cusped periodic wave, Right: Peaked periodic wave with $c=1$.
\par
\bigskip
\leftline{\bf 2.4. Remarks} \par
\medskip
\noindent 1. The asymptotic analysis of the general $N$-soliton solution is carried out following the method developed for the $N$-soliton solution of 
the Novikov equation [15].
Without entering into the detail, we summarize the result. The asymptotic form of the $N$-soliton solution is given by the superposition of $N$ cusp solitons each of which
is represented by the single cusp soliton (2.31).  The only difference is the phase shift caused by the collisions between cusp solitons. 
The explicit formula of the phase shift is derived from that of the $N$-soliton solution of the Novikov equation by means of a limiting procedure.
 To be more specific, let us introduce the scalings
$\Delta_i\rightarrow \epsilon\Delta_i, k \rightarrow  k_i/\epsilon$ in  formula (4.40) with  $\kappa=1/\sqrt{2}$ of [15] and then take the limit $\epsilon\rightarrow 0$.
The resulting expression  for the phase shift takes  the same form as that of the $N$-soliton solution of the Vakhnenko equation. See expression (40) of Ref. [11]. \par
\medskip
\noindent 2. Quite recently, the multicuspon solutions of equation (2.1) have been constructed by using the Darboux transformation [20].
The method of solution consists of solving the third-order ODE that comes from the spatial part of the Lax representation of equation (2.1). The parametric
solutions are represented by the  independent solutions of the ODE which would coincide with those obtained here by the direct method.
However, the proof of the equivalence of both solutions is not trivial and deserves a detailed analysis. \par
\medskip
\noindent 3. By means of the Galilean transformation $(x, t, u) \rightarrow (X, T, U)$ according to  $X=x+u_0t, T=t, U=u+u_0$ with  $u_0$ being a constant, equation (2.1) transforms to
$U_{XT}=(3/2)(U-u_0)-UU_{XX}-(1/4)U_X^2$. The solutions of this equation approaching  a constant $u_0$ as $|X| \rightarrow \infty$ are 
obtained simply from the present parametric solutions with zero boundary condition. \par
\bigskip
\noindent 4. The tau-function $f$ is given explicitly by (2.5), and $g$ is defined by (2.4). 
The explicit form of the tau-function $g$  may be obtained from  the tau-function corresponding to the $N$-soliton
solutions of the Novikov equation (Ref. [15] cited).  For the two-soliton tau-function given by (4.25) in [15],  one replaces $\xi_j$ and $k_j$ by $\xi_j+\pi{\rm i}$  and $k_j/\epsilon\ (j=1, 2)$, 
respectively and then takes the limit $\epsilon\rightarrow 0$. This gives the explicit two-soliton  tau-function g. This procedure can be applied to (3.1) in [15], yielding the explicit $N$-soliton tau-function $g$ for (2.4).  
As for the related BKP tau-functions,  the  paper by B-F. Feng {\it et al} [21] may be referred.  \par

\bigskip
\leftline{\bf 3. Parametric solutions of equation (1.6)}\par
\medskip
\noindent In this section, we present the parametric multisoliton solutions of equation (1.6).
First, we observe that equation (1.6) exhibits  two exact solutions, $u=-x/4+c_1t+c_2$ and $u=x/2+c_1t+c_2$, where $c_1$ and $c_2$ are
arbitrary constants. We shall show that soliton solutions of equation (1.6) asymptotically approach either these straight lines or $u=0$ as
$|x|\rightarrow \infty$. Hence, the zero boundary conditions $u\rightarrow 0, |x|\rightarrow\infty$ are not specified in advance. \par
  Here, we establish the following theorem. \par
  \medskip
\noindent {\bf Theorem 3.1.} {\it Equation (1.6) admits the parametric representation for the multisoliton solutions 
$$u={1\over 2}\left({\rm ln}\,{f^2\over g}\right)_{\tau}, \eqno(3.1a)$$
$$x=y-({\rm ln}\,f^4g)_\tau+y_0, \eqno(3.1b)$$
where the tau-functions $g=g(y, \tau)$ and $f=f(y, \tau)$ are given respectively by (2.4) and (2.5) in which the variable $\tau$ is 
replaced by $2\tau/3$.} \par
\bigskip
\leftline{\bf 3.1. Reciprocal transformation}\par
\medskip
\noindent Let
$$p=(1-2u_x)^{2\over 3}(1+4u_x)^{1\over 3}. \eqno(3.2)$$
Then, equation (1.6) can be written in the form of local conservation law
$$p_t+4(u^2p)_x=0. \eqno(3.3)$$
This allows us to introduce the reciprocal transformation $(x, t) \rightarrow (y, \tau)$ by
$$dy=pdx-4u^2pdt, \quad d\tau=dt. \eqno(3.4)$$
It follows from (3.4) that the variable $x=x(y, \tau)$ obeys the system of linear PDEs
$$x_y={1\over p}, \quad x_\tau=4u^2. \eqno(3.5)$$
The compatibility condition of above system of equations now gives a nonlinear PDE equivalent to equation (3.3)
under the reciprocal transformation (3.4). It reads
$$\left({1\over p}\right)_\tau=4(u^2)_y. \eqno(3.6)$$
Introduce the new variable $W$ by $W=pu_y$. Then, $u_x=pu_y=W$, and (3.2) is rewritten in terms of $W$ as
$$p^3=(1-2W)^2(1+4W). \eqno(3.7)$$
\par
Now,  equation (3.6) can be put into the form
$$p_\tau=-8puW. \eqno(3.8)$$
 We differentiate (3.7) by $\tau$ and use (3.8) to give
$$W_\tau={up^2\over \psi}, \quad \psi\equiv {1-2W\over p}. \eqno(3.9)$$
It follows from (3.7) and the definition of $\psi$ from (3.9) that
$${1\over \psi^2}={1+4W\over p}. \eqno(3.10)$$
Taking the logarithmic derivative of (3.10) with respect to $\tau$, using (3.8) and (3.9) to replace the
derivatives $p_\tau$ and $W_\tau$ and then differentiating the resultant equation by $y$, we obtain
 $({\rm ln}\,\psi)_{\tau y}=-2u_y=-2W/p$. It also follows from (3.9) and (3.10) that $\psi-\psi^{-2}=-6W/p$.
 Combining both expressions yields
 a nonlinear PDE for the variable $\psi$
$$({\rm ln}\,\psi)_{\tau y}={1\over 3}\,(\psi-\psi^{-2}). \eqno(3.11)$$
This is a form of the Tzitzeica equation [6]. \par
\bigskip
\leftline{\bf  3.2. Parametric representation of  solutions} \par
\medskip
\noindent Let $V=-\psi_{yy}/\psi$, or
$$\psi_{yy}+V\psi=0. \eqno(3.12)$$
Referring to  equation (3.11),  (3.12) gives a linear relation between $V$ and $\psi$ 
$$V_\tau=-\psi_y. \eqno(3.13)$$
Eliminating the variable $\psi$ from (3.12) and (3.13), we obtain  a single PDE for the variable $V$
$$VV_{\tau yy}-V_yV_{\tau y}+V^2V_\tau=0. \eqno(3.14)$$
Recall that equation (3.14) takes exactly the same form as equation (2.14). Keeping this fact in mind,
we introduce the dependent variable transformation $V=6({\rm ln}\,f)_{yy}$ for equation (3.13) and then integrate it with respect to $y$ to obtain
$$\psi=-6({\rm ln}\,f)_{\tau y}+\psi_0, \eqno(3.15)$$
where $\psi_0$ is an integration constant. To determine this constant, 
we impose the boundary condition $u\rightarrow 0$ as $x\rightarrow -\infty$. It then follows from the relations $W=u_x$ and  (3.7) that
${\rm lim}_{y\rightarrow -\infty}W=0,\ {\rm lim}_{y\rightarrow -\infty}\,p=1$.  These are used to obtain the limiting value of $\psi$, i.e., ${\rm lim}_{y\rightarrow -\infty}\psi={\rm lim}_{y\rightarrow -\infty}(1-2W)/p=1$. 
Thus, under the assumption ${\rm lim}_{y\rightarrow -\infty}({\rm ln}\,f)_{\tau y}=0$, we find $\psi_0=1$, so that (3.15) becomes
$$\psi=1-6({\rm ln}\,f)_{\tau y}. \eqno(3.16)$$
We can see from (2.14), (2.15), (2.22), (3.12),  (3.14) and (3.16) that the tau-function $f$ is given by (2.5)  with $\tau$ replaced by $2\tau/3$. 
We may impose the boundary condition $u\rightarrow 0$ as $x\rightarrow +\infty$ which yields the same limiting value 1 for $\psi$. 
Note that the boundary condition for $u$ is not compatible with the exact solutions of equation (1.6), i.e., $u=-x/4+c_1t+c_2$ and $u=x/2+c_1t+c_2$ for which
$W=-1/4$ and $W=1/2$, respectively. It turns out from (3.7) that $p=0$. As a result, the reciprocal transformation (3.4) becomes singular.
\par
It now follows from (3.5), (3.9) and the relation $W/p=u_y$ that $x_y=\psi+2u_y$. Substituting (3.16) into this expression and integrating it with respect to $y$, we obtain
$$x=y-6({\rm ln}\,f)_{\tau }+2u+c(\tau), \eqno(3.17)$$
where $c$ is an integration constant which depends generally on $\tau$.  Its $\tau$ dependence can be determined by the relation $x_\tau=u^2$ from (3.5) and the boundary conditions
${\rm lim}_{y\rightarrow -\infty}u=0, {\rm lim}_{y\rightarrow -\infty}({\rm ln}\,f)_{\tau \tau}=0$.  Consequently, one has  $c^\prime=0$, and  $c=y_0$(=const.).
Hence, (3.17) becomes
$$x=y-6({\rm ln}\,f)_{\tau }+2u+y_0. \eqno(3.18)$$
\par
The next step is to derive the expression of $x$ in terms of tau-functions. To begin with, we  define the tau-function $g$  by
$$g=f^2-3D_\tau D_yf\cdot f, \eqno(3.19)$$
which, taking into account (3.16), leads to an important relation $\psi=g/f^2$.
Since $f$ is given by (2.5) with $\tau$ replaced by $2\tau/3$,   the same recipe can be applied to  (2.24).  This yields the key
bilinear identity between the tau-functions $f$ and $g$
$$g^2-{3\over 2}\,D_\tau D_yg\cdot g=f^4. \eqno(3.20)$$
Dividing (3.20) by $g^2$ and using the relation $\psi^2=g^2/f^4$, one has
$${1\over \psi^2}={f^4\over g^2}=1-3({\rm ln}\,g)_{\tau y}. \eqno(3.21)$$
It follows from  (3.5),  (3.9) and (3.10) that
$$x_y={1\over p}={1\over 3}\left(2\psi+{1\over \psi^2}\right). \eqno(3.22)$$
If we introduce (3.16) and (3.21) into (3.22) and integrate the resultant equation with respect to $y$, we obtain
$$x=y-({\rm ln}\,f^4g)_\tau+y_0, \eqno(3.23)$$
where we have assumed the same boundary condition as that used in deriving (3.18).
Last, combining (3.18) and (3.23), we obtain the expression of $u$ in terms of the tau-functions $f$ and $g$
$$u={1\over 2}\left({\rm ln}\,{f^2\over g}\right)_\tau, \eqno(3.24)$$
which, together with (3.23), provides the parametric representation of the multisoliton solutions of equation (1.6). This completes  the proof of  Theorem 3.1.
\par
The expression of $u^2$ comes immediately from  (3.5) and (3.23). It reads
$$u^2=-{1\over 4}({\rm ln}\,f^4g)_{\tau\tau}. \eqno(3.25)$$
\medskip
Note that substitution of $u$ from (3.24) into (3.25) yields the bilinear identity between the tau-functions $f$ and $g$,
$4f_{\tau\tau}g-4f_\tau g_\tau+fg_{\tau\tau}=0$, which may
 be confirmed by a direct computation using the explicit forms of $f$ and $g$. 
 If we introduce a function $F$ by $f=\sqrt{F}$, then the above identity can be rewritten 
as  a trilinear form $gD_\tau^2F\cdot F +2FD_\tau^2F\cdot g=0$ for the tau-functions $F$ and $g$. 
This identity may be further decoupled to a set of bilinear identities $D_\tau^2F\cdot F=2hF$ and $D_\tau^2F\cdot g=-hg$ with $h$ being a
polynomial in ${\rm e}^{\xi_i}\ (i=1, 2, ..., N)$.
We recall that since $f$ is a pfaffian, $F$ admits a determinantal expression. See, for example Ref. [21].
 \par
\bigskip
\leftline{\bf 3.3.  Soliton solutions} \par
\medskip
\noindent The tau-functions for the one-soliton solution is given by  (2.4) and (2.5) with $\tau$ replaced by $2\tau/3$. They read
$$f=1-{\rm e}^\xi, \quad g=1+4{\rm e}^\xi+{\rm e}^{2\xi}, \quad \xi=ky+{\tau\over k}+\xi_0. \eqno(3.26)$$
Substitution of  these expressions into (3.1) yields the parametric representation of the one-soliton solution
$$u={3\over 2k}{{\rm coth}\,{\xi\over 2}\over \cosh\,\xi+2}, \eqno(3.27a)$$
$$X\equiv x+{t\over k^2}+x_0={\xi\over k}-{3\over 2k}{{\rm e}^{\xi\over 2}({\rm e}^{\xi}+3)\over \sinh\,{\xi\over 2}\,(\cosh\,\xi+2)}+y_0. \eqno(3.27b)$$
It follows from  $(3.27 a, b)$  by a direct computation that
$${du\over dX}=-{3(\cosh\,2\xi+2\,\cosh\,\xi+3)\over \cosh\,3\xi +6\, \cosh\,2\xi+39\, \cosh\,\xi+26}. \eqno(3.27c)$$
\begin{center}
\includegraphics[width=16cm]{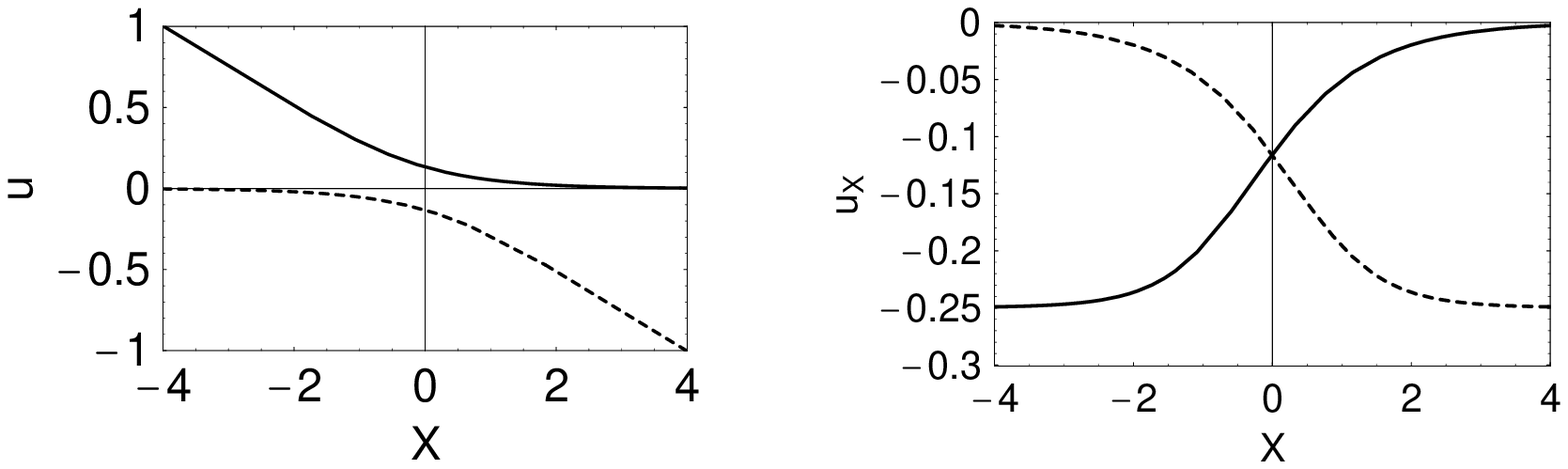}
\end{center}
\noindent{\bf Figure 2.}\ The profiles of an unbounded  solution  and its $X$ derivative  with the parameters $k=1.0, y_0=3$.  Left: $u$.\ Right: $u_X$.\par
\bigskip
The typical  profiles of $u$ and $u_X$ are depicted in figure 2 as a function of $X$.
The solution exhibits the different features depending on the range of the parameter $\xi$.  Specifically, the solution represented by the solid curve in figure  2 corresponds to $0<\xi<\infty$ whereas 
that of the dashed curve corresponds to $-\infty<\xi<0$. The former (latter) solution diverges when $X$ tends to $-\infty\ (+\infty) $, and asymptotically approaches a straight line $u=-X/4$ since $u\sim 1/(k\xi), 
X\sim-4/(k\xi)$  near $\xi=0$. We observe that $\xi=0$ is the zero of the tau-function $f$ from (3.26) which separates two branches of the solutions. 
One can show that $-1/4<u_X<0$ for any finite $X$. See $(3.27c)$ and remarks 3.4 below. Recall that  $u=-X/4$ is an exact solution of  equation (1.6).
It can be seen from figure 2 that for each curve, $u_X$ takes the form of a kink.
Note also that if $u(x, t)$ is a solution of equation (1.6), then $-u(-x,-t)$  satisfies equation (1.6) as well. This symmetry relation is  manifested clearly in
figure 2 in which the solution represented by the solid curve is mapped to that represented by the dashed curve by means of the  transformation $u(X) \rightarrow -u(-X)$.
\par
Another divergent solution is produced if one shifts the parameter $\xi_0$ as $\xi_0\rightarrow \xi_0+{\rm i}\pi$. Then, the tau-functions $f$ and $g$ from (3.26) become
$$f=1+{\rm e}^\xi, \quad g=1-4{\rm e}^\xi+{\rm e}^{2\xi}, \quad \xi=ky+{\tau\over k}+\xi_0. \eqno(3.28)$$
The parametric representation of the solution  corresponding to (3.27) is given by
$$u={3\over 2k}{{\rm tanh}\,{\xi\over 2}\over 2-\cosh\,\xi}, \eqno(3.29a)$$
$$X\equiv x+{t\over k^2}+x_0={\xi\over k}-{3\over 2k}{{\rm e}^{\xi\over 2}(3-{\rm e}^{\xi})\over \cosh\,{\xi\over 2}\,(2-\cosh\,\xi)}+y_0,\eqno(3.29b)$$
$${du\over dX}={3(\cosh\,2\xi-2\,\cosh\,\xi+3)\over \cosh\,3\xi -6\, \cosh\,2\xi+39\, \cosh\,\xi-26}. \eqno(3.29c)$$
\begin{center}
\includegraphics[width=16cm]{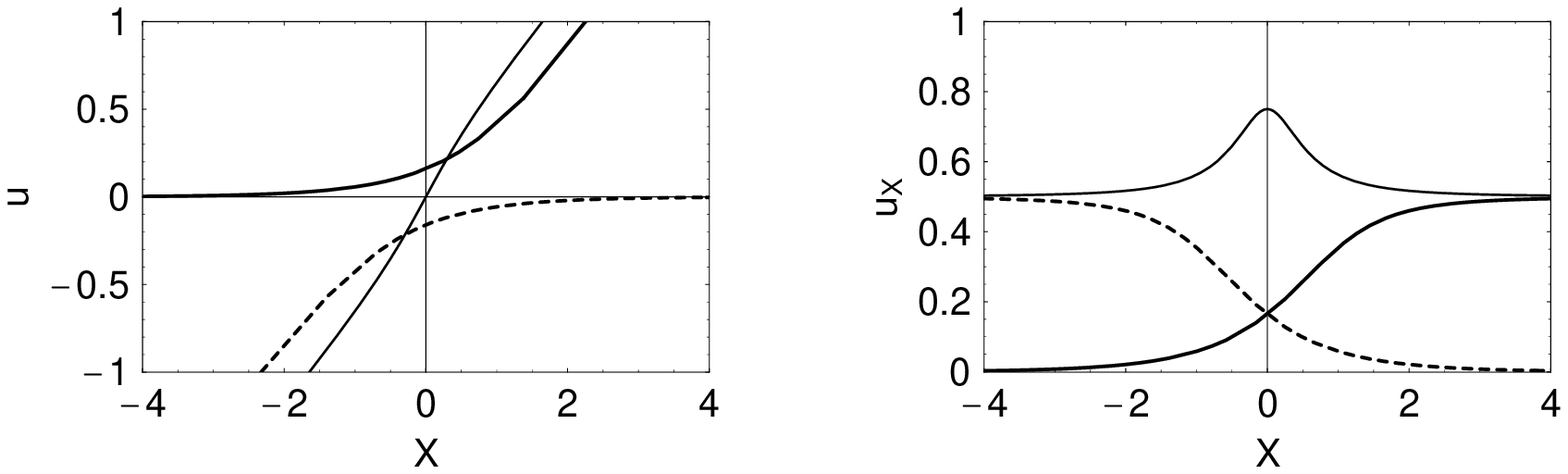}
\end{center}
\noindent {\bf Figure 3.}\ The profiles of an unbounded  solution  and its $X$ derivative  with the parameters $k=1.0, y_0=3$. 
Left: $u$. Right: $u_X$. \par
\bigskip
The  profiles of $u$ and $u_X$ for the same values of the parameters as those of figure 2  are shown in figure 3. The solution has three branches according to the range of  $\xi$.  The solid, thin solid and dashed curves
represent the solutions for $-\infty<\xi<-s, -s<\xi<s$ and $s<\xi<\infty$, respectively, where $s$ is a positive solution of the transcendental equation $\cosh\,\xi=2$, i.e., $s\simeq 1.317$.
This coincides with the positive zero of the tau-function $g$ from (3.28), i.e., $\xi={\rm ln}(2+\sqrt{3})$. 
When compared with figure 2, a new branch of the solution appears which is represented by the thin solid curve. The corresponding expression of  $u_X$  takes the form of  a bright soliton on a constant background.
Note  that the inequality $1/2<u_X<3/4$ holds for finite $X$, as will be shown in \S 3.4 remarks 1.  See the right panel of figure 3.
\par
Last, we briefly discuss the structure of the $N$-soliton solution. We assume that $f$ has $n$ real zeros with respect to $y$ at fixed $\tau$ and $g>0$ which reduce to (3.26) for $N=1$.
The opposite case corresponding to (3.28) can be dealt with in the same way.  Now, let the zeros of $f$ be $y_1<y_2<...<y_n$. Then, the solution would  have $n+1$ branches. Two of them exhibit
similar profiles to those represented by figure 2 whereas the remaining $n-1$ solutions look like the profile shown by the thin solid curve in figure 3.
Although numerical computations reveal that $n=N$, its rigorous verification deserves a detailed analysis. \par
\bigskip
\leftline{\bf 3.4. Remarks} \par
\medskip
\noindent 1. It follows from (3.9) and (3.10) that
$$u_y={W\over p}={1\over 6}\left({1\over\psi^2}-\psi\right). \eqno(3.30)$$
Using (3.22), (3.30) and the relation $\psi=g/f^2$, one has
$$u_x={u_y\over x_y}=-{1\over 4}\left(1-{{3\over 2}f^6\over g^3+{f^6\over 2}}\right). \eqno(3.31)$$
In the case of the tau-functions (3.26) for the one-soliton solution, one can easily check the inequalities $g>0$ and $g>f^2$ so that the relation (3.31) implies that $-1/4<u_x<0$.  
The explicit expression of $du/dX$ has been given by $(3.27c)$ from which one can confirm this inequality.
For the general $N$-soliton tau-functions with $k_i, \xi_{i0}\ (i=1, 2, ..., N)\in \mathbb{R}$, the above two inequalities hold 
thanks to the relation  (3.19),  provided that $D_\tau D_yf\cdot f<0$.  
\par
If the inequalities $f>0$ and $f^2>g$ hold for $f$ and $g$, we rewrite (3.31) as 
$$u_x={1\over 2}-{3\over 4}{g^3\over g^3+{f^6\over 2}}, \eqno(3.32)$$
and see that $0<u_x<1/2$ for $g>0$. This inequality is applied to  the  solid and  dashed curves in figure 3. 
If $g<0$ and $g^3+f^6/2>0$, on the other hand, then, $1/2<u_x<\gamma$, where $\gamma$ is a positive constant.  
 These conditions are indeed satisfied by the tau-functions (3.28).   One then finds that $\gamma=3/4$.  Actually, the inequality $1/2<u_X<3/4$ holds for the thin solid curve 
 in figure 3, as can be seen from  $(3.29c)$ for $-s<\xi<s,\ s=\ln(2+\sqrt{3})$.
 It turns out from the above discussion that while $u$ itself always diverges, its $x$ derivative $u_x$ is bounded and takes the form of either kink  or soliton, as already
 illustrated in figure 2 and figure 3.  
\par
\medskip
\noindent 2. Introducing the traveling wave ansatz $u=U(X), X=x-ct+x_0\ (c, x_0\in \mathbb{R})$ in  equation (1.6), one can recast it to a nonlinear ODE 
 $$\left(v-{1\over 2}\right)^2\left(v+{1\over 4}\right)={c^3d\over 16(4u^2-c)^3}, \eqno(3.33)$$
where $v=du/dX$ and $d (\in \mathbb{R})$ is an integration constant. In the case of $c<0$, an application of the theory of dynamical systems to  (3.33)
reveals that no bounded solutions exist. Specifically, for $d=1$, equation (3.33) can be integrated analytically to  yield the parametric solutions (3.27) and (3.29).
If $c>0$, on the other hand, then smooth periodic wave solutions are found to exist for the parameter $d$ in the interval $0<d<1$. 
The classification of  traveling periodic wave solutions of Eq. (3.33)  will be performed in a separate paper. 
\par
\bigskip
\leftline{\bf 4. Parametric solutions of equation (1.7)}\par
\medskip
\noindent We modify equation (1.7) by means of the linear transformation $(u, x, t)\rightarrow (U, X, T)$ according to 
$U={\beta\over\alpha}u+1, X={\sqrt{-\beta}\over\alpha}\,x+{\alpha\over\sqrt{-\beta}}\,t, T={\alpha\over\sqrt{-\beta}}\,t$. 
Then, we find that it is expressed in terms of the new variables as
$U_{XT}=U-1-U^2U_{XX}-UU_X^2$. Subsequently, we replace the variables $U, X, T$ by $u, {\rm i}x, -{\rm i}t$, respectively, and obtain the basic equation
that we consider here:
$$u_{xt}=u-1+u^2u_{xx}+uu_x^2. \eqno(4.1)$$
In this section, we solve equation (4.1) under the boundary condition $u\rightarrow 1, |x|\rightarrow \infty$ which corresponds to non-periodic solutions.  
\par
Here, we establish the following theorem. \par
\medskip
\noindent {\bf Theorem 4.1.} {\it Equation (4.1) admits the parametric representation for the multisoliton solutions 
$$u=1+{\rm i}\left({\rm ln}\,{\bar f\over g}\right)_\tau, \eqno(4.2a)$$
$$x=y-\tau-{\rm i}\,{\rm ln}\,{\bar f\over g}-({\rm ln}\,\bar fg)_\tau+y_0, \eqno(4.2b)$$
where  $\bar f$ and $g$ are tau-functions given by
$$ \bar f=\sum_{\mu=0,1}{\rm exp}\left[\sum_{i=1}^N\mu_i\left(\xi_i+d_i-{\pi\over 2}\,{\rm i}\right)
+\sum_{1\le i<j\le N}\mu_i\mu_j\gamma_{ij}\right], \eqno(4.3a)$$
$$ g=\sum_{\mu=0,1}{\rm exp}\left[\sum_{i=1}^N\mu_i\left(\xi_i-d_i+{\pi\over 2}\,{\rm i}\right)
+\sum_{1\le i<j\le N}\mu_i\mu_j\gamma_{ij}\right], \eqno(4.3b)$$
with
$$\xi_i=p_iy+{1\over p_i}\,\tau+\xi_{i0}, \quad (i=1, 2, ..., N),\eqno(4.3c)$$
$${\rm e}^{d_i}=\sqrt{1+{\rm i} p_i\over 1-{\rm i}p_i}, \quad (i=1, 2, ..., N),\eqno(4.3d)$$
$${\rm e}^{\gamma_{ij}}=\left({p_i-p_j\over p_i+p_j}\right)^2, \quad (i, j=1, 2, ..., N; i\not= j). \eqno(4.3e)$$
Here, $p_i$ and $\xi_{i0}$ are arbitrary complex parameters.}
\par
 If one puts $p_i=\tan\,\theta_i$, then the expression of $d_i$ from $(4.3d)$ simplifies to
$d_i={\rm i}\theta_i$.  This parametrization is found to be very useful in constructing cusp and breather solutions.
 \par
\bigskip
\leftline{\bf 4.1. Reciprocal transformation}\par
\medskip
\noindent In accordance with the observation that the form of equation (4.1) is quite similar to equation (1.4), we apply 
the same type of the reciprocal transformation\ $(x, t)\rightarrow (y, \tau)$ as used for equation (1.4) [10]
$$dy=rdx+ru^2dt, \quad d\tau=dt, \eqno(4.4)$$
where $r=r(x, t)$ is an unknown function subjected  to the  boundary condition $r\rightarrow 1, |x|\rightarrow \infty$.
It leads to the linear system of PDEs for $x=x(y, \tau)$
$$x_y={1\over r}, \quad x_\tau=-u^2, \eqno(4.5)$$
whose compatibility condition  yields
$$r_\tau=2r^2uu_y. \eqno(4.6)$$
Under the transformation (4.4), equation (4.1)  can be written in the form
$$ru_{\tau y}=u-1-r^2uu_y^2. \eqno(4.7)$$
\par
Now, we introduce the new variable $w$ by
$$w=ru_y, \eqno(4.8)$$
to transform equations (4.6) and (4.7) to
$$r_\tau=2ruw, \eqno(4.9)$$
$$w_\tau-{r_\tau\over r}\,w=u-1-uw^2, \eqno(4.10)$$
respectively.  Note from (4.5) and (4.8) that $u_x=w$. If we eliminate the variable $r_\tau$ from (4.9) and (4.10), we obtain the expression of $u$ in terms of $w$
$$u={w_\tau+1\over w^2+1}. \eqno(4.11)$$
We further introduce the  variable $\phi$ according to
$$w=\tan\,{\phi\over 2}. \eqno(4.12)$$
With this variable, (4.11) can be written as
$$u={1\over 2}\,\phi_\tau+\cos^2{\phi\over 2}, \eqno(4.13)$$
and $r$ from (4.8) recasts to
$$r={w\over u_y}={2\tan\,{\phi\over 2}\over \phi_{\tau y}-\sin\,\phi\, \phi_y}. \eqno(4.14)$$
Substituting (4.13) and (4.14) into (4.6), we obtain a single nonlinear PDE for $\phi$
$$\phi_{\tau\tau y}-{1\over\tan\,\phi}\,\phi_\tau\phi_{\tau y}-\sin^2\phi\,\phi_y=0. \eqno(4.15)$$
Remarkably, equation (4.15) can be integrated once with respect to $y$.  Indeed, if we multiply (4.15) by an integrating factor $\phi_{\tau y}/2$ and perform the
 integration, we obtain
 $$\left({\phi_{\tau y}\over \sin\,\phi}\right)^2-\phi_y^2=c(\tau), \eqno(4.16)$$
 where $c$ is an integration constant.  Now, ${\rm lim}_{|y|\rightarrow\infty}\,w=0$ by virtue of the boundary condition for $u$, and hence  
 ${\rm lim}_{|y|\rightarrow\infty}\,\phi=0 \ ({\rm mod}\ 2\pi)$ by (4.12).
It follows from (4.14) that ${\rm lim}_{|y|\rightarrow\infty}\,r=1={\rm lim}_{|y|\rightarrow\infty}\,\phi/\phi_{\tau y}$. 
Consequently, ${\rm lim}_{|y|\rightarrow\infty}\,\phi_{\tau y}/\sin\,\phi
={\rm lim}_{|y|\rightarrow\infty}\,\phi_{\tau y}/\phi=1$. Taking the limit $|y|\rightarrow\infty$ in (4.16) and using this limiting value, we  can determine  $c=1$, and  
finally equation (4.16) becomes
 $$\left({\phi_{\tau y}\over \sin\,\phi}\right)^2=1+\phi_y^2. \eqno(4.17)$$
The square root of equation (4.17) gives rise to the modified sG equation 
$${\phi_{\tau y}\over \sqrt{1+\phi_y^2}}=\sin\, \phi, \eqno(4.18)$$
where the plus sign has been chosen without loss of generality.  \par
\bigskip
\leftline{\bf  4.2.  Parametric representation of  solutions} \par
\medskip
\noindent The modified sG equation (4.18) is a completely integrable PDE, and its multisoliton solutions have been obtained 
in constructing solutions of the generalized sG equation [22]. 
The following proposition provides the parametric $N$-soliton solution. \par
\medskip
\noindent {\bf Proposition 4.1} {\it The parametric representation of the $N$-soliton solution of Eq. (4.18) is given by
$$\phi={\rm i}\,{\rm ln}\,{\bar f\bar g\over fg}, \eqno(4.19a)$$
$$x=y-\tau+{\rm i}\,{\rm ln}\,{\bar g g\over \bar f f}+y_0, \eqno(4.19b)$$
where the tau-functions $\bar f$ and $g$ are given respectively by $(4.3a)$ and $(4.3b)$, and $f$ and $\bar g$  read in the form
$$ f=\sum_{\mu=0,1}{\rm exp}\left[\sum_{i=1}^N\mu_i\left(\xi_i+d_i+{\pi\over 2}\,{\rm i}\right)
+\sum_{1\le i<j\le N}\mu_i\mu_j\gamma_{ij}\right], \eqno(4.20a)$$
$$ \bar g=\sum_{\mu=0,1}{\rm exp}\left[\sum_{i=1}^N\mu_i\left(\xi_i-d_i-{\pi\over 2}\,{\rm i}\right)
+\sum_{1\le i<j\le N}\mu_i\mu_j\gamma_{ij}\right]. \eqno(4.20b)$$
}
\par
It has been shown that the tau-functions $f, \bar f, g$ and $\bar g$ satisfy the bilinear equations [22]:
$$D_\tau D_yf\cdot f={1\over 2}\,(f^2-{\bar f}^2),\quad D_\tau D_y\bar f\cdot \bar f={1\over 2}\,({\bar f}^2-f^2),    \eqno(4.21a)$$
$$D_\tau D_yg\cdot g={1\over 2}\,(g^2-{\bar g}^2),\quad D_\tau D_y\bar g\cdot \bar g={1\over 2}\,({\bar g}^2-g^2),    \eqno(4.21b)$$
$${\rm i}D_y f\cdot \bar g={1\over 2}\,(f\bar g-\bar f g),\quad   {\rm i}D_y\bar f\cdot g={1\over 2}\,(\bar f g- f \bar g),  \eqno(4.22)$$
$${\rm i}D_\tau f\cdot g=-{1\over 2}\,(fg-\bar f\bar g),\quad  {\rm i}D_\tau \bar f\cdot \bar g=-{1\over 2}\,(\bar f\bar g- fg).  \eqno(4.23)$$
\par
\medskip
Let us now prove Theorem 4.1. First,  we combine (4.14) with (4.18) to eliminate the variable $\phi_{\tau y}$, giving
$$x_y={1\over r}=\left(\sqrt{1+\phi_y^2}-\phi_y\right)\,\cos^2{\phi\over 2}. \eqno(4.24)$$
We use $(4.19a)$ and (4.22) to compute $\phi_y$ and obtain
$$\phi_y=-{\rm i}\,{D_y f\cdot \bar g\over f\bar g}+{\rm i}\,{D_y \bar f\cdot g\over \bar fg}={(\bar fg)^2-(f\bar g)^2\over 2\bar f\bar g fg}. \eqno(4.25)$$
Substitution of  $(4.19a)$ and (4.25) into (4.24) gives 
$$x_y={1\over 4}\,\left({f\over \bar f}+{\bar g\over g}\right)^2. \eqno(4.26)$$
It follows from $(4.21a)$ and $(4.21b)$ that
$$({\rm ln}\,\bar fg)_{\tau y}={D_\tau D_y\bar f\cdot \bar f\over 2\bar f^2}+{D_\tau D_y g\cdot g\over 2g^2}=1-{1\over 2}\left(1-{f\bar g\over \bar f g}\right)
-{1\over 4}\,\left({f\over \bar f}+{\bar g\over g}\right)^2, \eqno(4.27)$$
and from (4.22) that
$${\rm i}\,\left({\rm ln}\,{\bar f\over g}\right)_y={\rm i}\, {D_y \bar f\cdot g\over \bar fg}={1\over 2}\left(1-{f\bar g\over \bar f g}\right). \eqno(4.28)$$
In view of (4.27) and (4.28), (4.26) is expressed as
$$x_y=1-{\rm i}\,\left({\rm ln}\,{\bar f\over g}\right)_y-({\rm ln}\,\bar fg)_{\tau y}, \eqno(4.29)$$
which can be integrated with respect to $y$ to give
$$x=y-{\rm i}\,{\rm ln}\,{\bar f\over g}-({\rm ln}\,\bar fg)_{\tau }+c(\tau). \eqno(4.30)$$
The integration constant $c$ is determined from the boundary condition ${\rm lim}_{|y|\rightarrow\infty}u=1$. Actually,
${\rm lim}_{|y|\rightarrow\infty} x_\tau={\rm lim}_{|y|\rightarrow\infty}(-u^2)=-1$ and 
${\rm lim}_{|y|\rightarrow\infty}\left({\rm ln}\,{\bar f\over g}\right)_\tau=0, {\rm lim}_{|y|\rightarrow\infty}\left({\rm ln}\,{\bar f\over g}\right)_{\tau\tau}=0$.
Referring to these limiting values, one has $c^\prime=-1$ so that $c=-\tau+y_0$ which, substituted in (4.30), yields the parametric representation $(4.2b)$ for $x$. \par
To derive the parametric representation of $u$, we introduce $(4.19a)$ into (4.13) to obtain
$$u={{\rm i}\over 2}\,\left({\rm ln}\,{\bar f\bar g\over fg}\right)_\tau+{1\over 2}\left\{1+{1\over 2}\left({\bar f\bar g\over fg}+{fg\over\bar f\bar g}\right)\right\}. \eqno(4.31)$$
On the other hand, it follows from (4.23) that
$${\rm i}\left({\rm ln}\,{f\over g}\right)_\tau=-{1\over 2}\,\left(1-{\bar f\bar g\over fg}\right),\quad 
{\rm i}\left({\rm ln}\,{\bar f
\over \bar g}\right)_\tau=-{1\over 2}\,\left(1-{ fg\over \bar f\bar g}\right). \eqno(4.32)$$
Adding both equations placed in (4.32) gives
$${1\over 2}\,\left({\bar f\bar g\over fg}+{ fg\over \bar f\bar g}\right)=
1+{\rm i}\left({\rm ln}\,{\bar ff\over \bar gg}\right)_\tau. \eqno(4.33)$$
Last, substituting the left-hand side of (4.33) into (4.31), we obtain
$$u=1+{\rm i}\,\left({\rm ln}\,{\bar f\over g}\right)_\tau, \eqno(4.34)$$
which is just $(4.2a)$. This completes the proof of Theorem 4.1. \par
The  solutions given by (4.2) become complex-valued functions for complex parameters $p_i$ and $\xi_{i0}$.
If one imposes the conditions $\bar f=g^*, \bar g=f^*$ for the tau-functions with the asterisk being the complex conjugate, then,   (4.2)
gives rise to real solutions characterized by a single tau-function $g$. It reads
$$u=1+{\rm i}\left({\rm ln}\,{g^*\over g}\right)_\tau, \eqno(4.35a)$$
$$x=y-\tau-{\rm i}\,{\rm ln}\,{g^*\over g}-({\rm ln}\,g^*g)_\tau+y_0. \eqno(4.35b)$$
The conditions mentioned above for the tau-functions are realized if one takes the real values for the parameters $p_i$ and $\xi_{i0}$ which would  lead to
cusp soliton solutions, as will be exemplified below.
 Another choice is possible by putting $N=2M$ and $p_{2i-1}=p_{2i}^*, \xi_{2i-1}=\xi_{2i}^*\ (i=1, 2, ..., M)$, where $M$ is a
positive integer.   The resulting solutions  would be shown to take the form of breather solutions. \par
\bigskip
\leftline{\bf 4.3. Soliton and periodic solutions} \par
\medskip
\leftline{\bf 4.3.1. Cusp soliton} \par
\noindent The tau-function $g$ for the one-soliton solution is given by $(4.3b)$ with $N=1$. It can be written  in the form
$$g=1+{\rm i}\,{\rm e}^{\xi-d}, \quad {\rm e}^{d}=\sqrt{1+{\rm i}p\over 1-{\rm i}p}, \quad \xi=py+{1\over p}\,\tau+\xi_0, \eqno(4.36)$$
where $p$ and $\xi_0$ are real parameters.  If we put $p=\tan\,\theta\ (0<\theta <{\pi/ 2})$, then $d={\rm i}\theta$ and 
$g=1+{\rm i}{\rm e}^{\xi-{\rm i}\theta}$. Introducing $g$ into (4.35) yields  a one-soliton solution
$$u=1+{\cos\,\theta\over\tan\,\theta}{1\over \cosh\,\xi+\sin\,\theta}, \eqno(4.37a)$$
$$X\equiv x+{1\over \sin^2\theta}\,t+x_0={\xi\over \tan\,\theta}-{1\over \tan\,\theta}{\sinh\,\xi\over \cosh\,\xi+\sin\,\theta}
-2\,\tan^{-1}\left({\cos\,\theta\over 1+\sin\,\theta}\,\tanh\,{\xi\over 2}\right). \eqno(4.37b)$$
Here, the constant  $y_0$ has been  chosen such that $\xi=0$ corresponds to $X=0$.
Although the expression (4.37) represents a bounded solution, it has a cusp at the crest. To see this, we compute the $X$ derivative of $u$ to obtain
$$u_X={u_\xi\over X_\xi}=-{\cos\,\theta\over\sinh\,\xi}. \eqno(4.37c)$$
 On the other hand, the coordinate $X$ from $(4.37b)$ has an expansion $X=\kappa\,\xi^3+O(\xi^5)$  near $\xi=0$, where
$\kappa={\cos\,\theta+\cot\,\theta\over 3(1+\sin\,\theta)^3}$. 
Thus, $u_X\sim -\kappa^{1/3}\,\cos \,\theta \,X^{-1/3}$, implying that
${\rm lim}_{X\rightarrow \pm 0}u_X=\mp\infty$. \par
The profile of the cusp soliton is plotted  as a function of $X$ on the left panel of figure 4. 
It represents a singular pulse propagating to the left at a constant velocity $1/\sin^2\,\theta$  on the coordinate system at rest.
\begin{center}
\includegraphics[width=16cm]{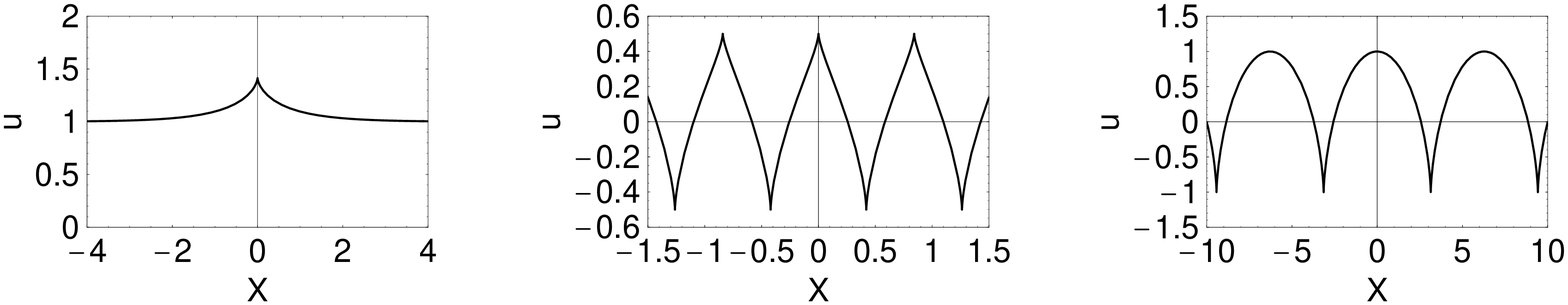}
\end{center}
\noindent {\bf Figure 4.}\ The profiles of the cusp soliton and cusped periodic wave  solutions.   
Left:  Cusp soliton with the parameter $\theta=\pi/4$. Middle: Cusped periodic wave with the parameter $\theta=\cosh^{-1}2$.
Right: Cycloid.
\par
\medskip
\leftline{\bf 4.3.2.  Cusped periodic wave} \par
\noindent The cusped periodic wave  solution is obtained formally from the cusp soliton solution (4.37) if one replaces the parameters
$\xi$ and $\theta$ by ${\rm i}\xi$ and $\pi/2+{\rm i}\theta$, respectively. The parametric solution (4.37) then becomes
$$u=1-{\sinh\,\theta\over\coth\,\theta}{1\over \cos\,\xi+\cosh\,\theta}, \eqno(4.38a)$$
$$X\equiv x+{1\over \cosh^2\theta}\,t+x_0={\xi\over \coth\,\theta}-{1\over \coth\,\theta}{\sin\,\xi\over \cos\,\xi+\cosh\,\theta}
-2\,\tan^{-1}\left({\sinh\,\theta\over 1+\cosh\,\theta}\,\tan\,{\xi\over 2}\right), \eqno(4.38b)$$
for $\xi\in (-\pi, \pi)$. The solution (4.38) can be periodically continued beyond the interval $(-\lambda/2, \lambda/2)$ with $\lambda=2\pi(1-\tanh\,\theta)$.
It turns out that  (4.38) represents a cusped periodic wave with the amplitude $1/\cosh\,\theta$ and the period $\lambda$.
Its typical profile is depicted on the middle panel of figure 4. Since $u_X=\sinh\,\theta/\sin\,\xi$, the cusp appears both
at the crest  and at the trough in marked contrast to  
the periodic traveling wave solution of equation (1.5) shown on the middle panel of figure 1 in which the cusp appears only at the crest of the wave.
\par
Last, we demonstrate that an intriguing result is obtained from a special limit of (4.38).  To be more specific, let $\xi=\pi +\theta\,\cot(\phi/2)\ (\phi\in \mathbb{R})$, and
take the limit $\theta \rightarrow 0$. Then, (4.38) reduces to
$$u=\cos\,\phi, \quad X\equiv x+t+x_0=\phi + \sin\,\phi. \eqno(4.39)$$
The profile of (4.39) is plotted on the right panel of figure 4.
It follows from (4.39) that $u_X=-\tan(\phi/2)$, which determines the position of the cusp, i.e., $X=\pi\ ({\rm mod}\ 2\pi)$.
The parametric solution (4.39) represents the cycloid which is familiar to us in geometry. 
It satisfies an  ODE, $(du/dX)^2=(1-u)/(1+u)$.  
Actually, if we seek the traveling wave solution of the form $u=u(X), X=x-ct+x_0\ (c, x_0\in \mathbb{R})$, then we  see that  Eq. (4.1) can be recast to
a nonlinear ODE, $(du/dX)^2=-(u^2-2u+d)/(u^2+c)\ (d\in \mathbb{R})$.  In the case of $c=-1$ and $d=1$, this equation simplifies to an ODE mentioned above.
Recall that the conventional representation of the cycloid is  given by $x=a(t-\sin\,t), y=a(1-\cos\,t),\ (a>0, t\in \mathbb{R})$ in the
$(x, y)$ plane, and 
the transformation of (4.39) to this form can be made simply by shifting the variables  $u$, $X$ and $\phi$  by  $u-1$, $X+\pi$ and $\phi+\pi$, respectively.
It is interesting to observe that the expression $-u$ from (4.39) coincides with the surface profile of Gerstner's trochoidal wave which has been derived more than
200 years ago in the theory of deep gravity waves [23, 24].
Last, we point out that smooth periodic wave solutions exist if $c>0$ and  $d<1$ in the above ODE.
\par
\medskip
\leftline{\bf 4.3.3.  Breather} \par
\noindent The breather solutions are obtained following the recipe described in the last sentence in section 4.2. 
To show this, we find it convenient to introduce the new parameters according to the relations 
$$\xi_j=\eta_j+{\rm i}\chi_j,\quad \xi_{j0}=\lambda_j+{\rm i}\mu_j,\quad  p_j=\tan\,\theta_j=a_j+{\rm i}b_j,\quad  \theta_j=\alpha_j+{\rm i}\beta_j,\quad (j=1, 2, ..., N). \eqno(4.40) $$
The parameters $a_j$ and $b_j$ are expressed in terms of $\alpha_j$ and $\beta_j$ as
$$a_j={{1\over 2}\,\sin\,2\alpha_j \over \cosh^2\beta_j-\sin^2\alpha_j}, \quad b_j={{1\over 2}\,\sinh\,2\beta_j \over \cosh^2\beta_j-\sin^2\alpha_j}, \quad (j=1, 2, ..., N). \eqno(4.41)$$
It turns out from (4.41) that $a_j/b_j=\sin\,2\alpha_j/\sinh\, 2\beta_j$. This quantity  characterizes the feature of the solution. 
\par
\begin{center}
\includegraphics[width=16cm]{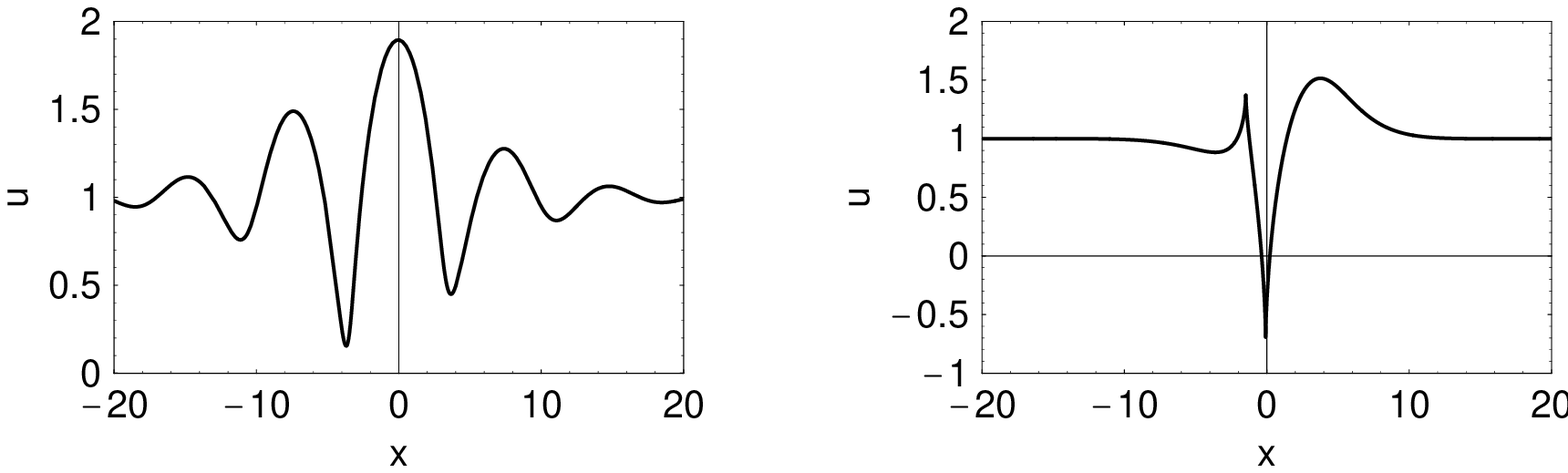}
\end{center}
\noindent{\bf Figure 5.}\ The profiles of the  breather solutions.  Left: The smooth breather with the parameters 
$p_1=p_2^*=\tan(\pi/6+{\rm i}), \xi_{10}=\xi_{20}^*=0$. Right: The singular breather with the parameters 
$p_1=p_2^*=\tan(\pi/6+0.2\,{\rm i}), \xi_{10}=\xi_{20}^*=0$.
\par
\medskip
Here, we consider the single breather solution. The corresponding tau-function is given by $(4.3b)$ with $N=2$.  We put
$\xi_1=\xi_2^*=\eta+{\rm i}\chi, p_1=p_2^*=\tan\,\theta=a+{\rm i}b, \theta=\alpha+{\rm i}\beta,  \xi_{10}=\xi_{20}^*=\lambda+{\rm i}\mu,  \delta=(b/a)^2$. 
Then, the tau-function $f$ and $g$ can be
expressed in the form
$$f=1+{\rm i}\,{\rm e}^{\eta-\beta+{\rm i}(\chi+\alpha)}+{\rm i}\,{\rm e}^{\eta+\beta-{\rm i}(\chi-\alpha)}+\delta\,{\rm e}^{2\eta+2{\rm i}\alpha}, \eqno(4.42a)$$
$$g=1+{\rm i}\,{\rm e}^{\eta+\beta+{\rm i}(\chi-\alpha)}+{\rm i}\,{\rm e}^{\eta-\beta-{\rm i}(\chi+\alpha)}+\delta\,{\rm e}^{2\eta-2{\rm i}\alpha}, \eqno(4.42b)$$
with
$$\eta=ay+{a\over a^2+b^2}\,\tau+\lambda, \quad \chi=by-{b\over a^2+b^2}\,\tau+\mu. \eqno(4.42c)$$
\par
The parametric representation of the solution is given by (4.35) with the tau-function $g$ from $(4.42b)$. The explicit form of the solution is, however too complicated to
write down here.
The profile of the smooth breather solution at $t=0$ is depicted on the left panel of figure 5 as a function of $x$ for the specified values of the parameters. 
It represents an oscillating pulse on a constant background. The smoothness of the solution depends on the parameter $1/\sqrt{\delta}=a/b \ (a>0, b>0)$.  
A detailed analysis reveals that the smooth solutions are produced if the condition $a/b<\cos\,\alpha/\cosh\,\beta\ (0<\alpha<\pi/2, \beta>0)$ is satisfied.
See remarks 4.4 below.
In the present example, $a/b=0.239, \cos\,\alpha/\cosh\,\beta=0.561\  (\alpha=\pi/6, \beta=1.0)$.
As for the analogous conditions for the smooth breather solutions of the SP and modified SP equations, one can refer to Refs. [10, 25].
\par
If the value of the parameter $a/b$ exceeds certain critical value, then  the singularities of the solutions appear in the form of cusps. 
The amplitude of the solution is, however finite since   $g\not=0 $ (or $g^*g>0$) for arbitrary values of $\eta$ and $\chi$, as can be confirmed by 
 using $(4.42b)$. 
The formation of  cusps is exemplified on the right panel of figure 5 for the specified values of the parameters. 
In this example, $a/b=2.11$ and $\cos\,\alpha/\cosh\,\beta=0.849$, so that the condition for the smoothness is violated.
\par
\bigskip
\leftline{\bf 4.4. Remarks} \par
\medskip
\noindent 1. The smooth breather solutions would be produced if the conditions $g\not=0$ and ${\rm Re}(fg)\not=0$ are satisfied   simultaneously  for arbitrary values of $y$ and $\tau$.
Referring to $(4.35a)$, the former condition assures that the solution is finite  in space and time,
whereas the latter one implies that the mapping (4.4) is one-to-one so that the solution becomes a single-valued function of $x$ at any fixed $t$. 
Actually,  in view of (4.26) with $\bar f=g^*$ and $\bar g=f^*$, one obtains $x_y=({\rm Re}\,fg)^2/(g^*g)^2$. Hence, $x_y>0$ if ${\rm Re}\,fg\not=0$ and $g^*g>0$.
It also follows from (4.12), $(4.19a)$ and the relation $u_x=w$ that $u_x={\rm Im}\,fg/{\rm Re}\, fg$. Thus, if ${\rm Re}\,fg\not=0$, then the singularities $u_x\rightarrow \pm\infty$
never occur in the solution. In the case of one-breather solutions, the condition ${\rm Re}\, fg>0$  is equivalent to imposing the inequality  $\sin\,\alpha<\sinh\,\beta\ (0<\alpha<\pi/2, \beta>0).$
In view of (4.41), this inequality can be rewritten  in the form $0<a/b<\cos\,\alpha/\cosh\,\beta$ which turns out to be a criterion for the existence of smooth breathers.
 \par
\medskip
\noindent 2. The modified SP equation (1.4)
 has been proposed as an integrable generalization of the SP equation [9]. Subsequently, an integrable multi-component generalization of the modified SP
equation was introduced as well as its multisoliton solutions [10].  The parametric representation of the $N$-soliton solution of equation (1.4) can be written in the form [10]
$$u={\rm i}\left({\rm ln}\,{\bar f_{sG}\over f_{sG}}\right)_\tau, \quad x=y-({\rm ln}\,\bar f_{sG}f_{sG})_\tau+y_0, \eqno(4.43)$$
where $f_{sG}$ and $\bar f_{sG}$ are tau-functions representing the $N$-soliton solution  of the sG equation $\phi_{\tau y}=\sin\,\phi$, and
given explicitly by $g$ from $(4.3b)$ and $\bar f$ from $(4.3a)$ with $d_i=0$, respectively.
The structure of the solutions is quite similar to that of equation (4.1). The only difference  is the parametric form of $x$
coming, in particular, from the term ${\rm i}\,{\rm ln}(\bar f/g)$ in $(4.2b)$.
 \par
\medskip
\noindent 3. If one carries out the scalings $u\rightarrow -u, x\rightarrow -{\rm i}x, t\rightarrow {\rm i}t$, equation (4.1) reduces to
$$u_{xt}=u+1-u^2u_{xx}-uu_x^2. \eqno(4.44)$$
The above equation has been considered in [13], where the parametric solutions are shown to be constructed in terms of solutions of the sG equation. 
The parametric solutions of equation (4.44) are obtained simply by applying the scalings $y\rightarrow -{\rm i}y, \tau\rightarrow {\rm i}\tau, y_0\rightarrow -{\rm i}y_0$
in addition to the scalings  mentioned above for $u, x$ and $t$ to (4.2).  This gives
$$u=-1-\left({\rm ln}\,{\bar f\over g}\right)_\tau, \eqno(4.45a)$$
$$x=y+\tau+{\rm ln}\,{\bar f\over g}-({\rm ln}\,\bar fg)_\tau+y_0, \eqno(4.45b)$$
where the tau-functions $\bar f$ and $g$  are given by (4.3) with the parameters $\theta_i$ replaced by $\pi/2+{\rm i}\theta_i\ (i=1, 2, ..., N)$.
These tau-functions  are real functions for the real soliton parameters $\theta_i$ and $\xi_{i0}$. Consequently,  
the parametric representation (4.45) gives rise to real solutions. Actually, 
in the  one-soliton case, the tau-functions $f$ and $g$ are expressed as
$$\bar f=1+{\rm e}^{\xi-\theta}, \quad g=1+{\rm e}^{\xi+\theta},\quad \xi=py+{\tau\over p}+\xi_0, \quad p=1/\tanh\,\theta, \eqno(4.46)$$
and the parametric one-soliton solution takes the form
$$u=-1+{\sinh\,\theta\over \coth\,\theta}\,{1\over \cosh\,\xi+\cosh\,\theta}, \eqno(4.47a)$$
$$X=x-{1\over \cosh^2\theta}\,t+x_0=\tanh\,\theta\,\xi-{\tanh\,\theta\,\sinh\,\xi\over\cosh\,\xi+\cosh\,\theta}
+{\rm ln}\,{1-\tanh\,{\theta\over 2}\,\tanh\,{\xi\over 2}\over 1+\tanh\,{\theta\over 2}\,\tanh\,{\xi\over 2}}, \eqno(4.47b)$$
with $du/dX=-\sinh\,\theta/\sinh\,\xi$.
 This represents a cusp soliton on a constant background $u=-1$ moving to the right at a constant velocity $1/\cosh^2\theta$. 
  The general $N$-cusp soliton solutions as well as the breather solutions can be obtained 
 in the same way following the procedure developed in section 4.3. 
\par
\bigskip
\leftline{\bf 5. Concluding remarks} \par
\medskip
\noindent In this paper, we have developed a systematic method for solving the integrable generalized short pulse equations. 
We have presented parametric solutions for three novel equations  by means of a direct method combined with the
reciprocal transformations. The solutions obtained include cusp solitons, unbounded solutions and breathers 
as well as singular periodic traveling wave solutions. These exhibit several new features. Of particular interest is that
no smooth and bounded solutions were found for the non-periodic case except for the breather solutions of equation (4.1).
The results of the present paper open up a number of interesting problems, some of which are listed below. 
 \par
 \medskip
 \noindent1. The  periodic   solutions presented in this paper belong to a special class of solutions. 
 The classifications of  traveling wave solutions  of equations (1.5), (1.6) and (4.1)  are important issues to be considered in detail.
Furthermore, the integrability of these equations suggests the existence of the general multiphase solutions as in the cases of the integrable
 PDEs such as the Korteweg-de Vries, sG and nonlinear Schr\"odinger (NLS) equations. Various exact methods are now available to construct  periodic solutions
 mentioned above [26, 27].
 \par
 \medskip
 \noindent 2. The Cauchy problems of the equations are more challenging than constructing particular solutions.  Specifically, 
 the  mathematical problems associated with the existence and uniqueness of  solutions as  well as  the well-posedness 
in  appropriate function spaces
  are worth studying  for both non-periodic and periodic boundary conditions [28]. The IST is
applicable to three PDEs since the corresponding Lax pairs are now at hand.   
The rigorous mathematical analysis of  the stability of traveling wave solutions is also an important issue  to be addressed in a future work.
\par
\medskip
\noindent 3. The SP and modified SP equations have been extended to integrable multi-component systems and their parametric solutions have been obtained [10, 29].  One can expect that
the single component PDEs considered here have integrable multi-component analogs. The two-component case is particularly important since it may be written as a single equation for the
complex variable with either  the focusing or defocusing nonlinearities, as in the case of the NLS equation.  \par
\medskip
\noindent 4. While the SP equation  has been derived from the physical system, other six equations still remain mathematical models.  
Therefore, it seems to be important to capture the physical phenomena described by these equations. 
In the context of surface gravity waves, a few works along this line are found, for example, in  [12, 18, 30]. 
\par
\bigskip
\noindent{\bf Acknowledgement}\par
\bigskip
\noindent The author appreciated critical review comments from two anonymous reviewers, which greatly improve an earlier draft of the manuscript. 
\par

  \newpage

\baselineskip=5.6mm
\noindent {\bf References}\par
\begin{enumerate}[{[1]}]{\baselineskip=5.5mm  \itemsep=0mm \parsep=0mm}
\item Sakovich A and  Sakovich S 2005  The short pulse equation is integrable  {\it J. Phys. Soc. Jpn.} {\bf 74} 239-241   \vspace{-1.5mm}
\item Rabelo M L 1989 On equations which describe pseudospherical surfaces  {\it Stud. Appl. Math.} {\bf 81} 221-248   \vspace{-1.5mm}
\item Beals R, Rabelo M  and  Tenenblat K 1989  B\"acklund transformation and inverse scattering solutions for some pseudospherical surface equations {\it Stud. Appl. Math. }{\bf 81} 125-151  \vspace{-1.5mm}
\item Sch\"afer T  and  Wayne C E 2004 Propagation of ultra-short optical pulses in cubic nonlinear media {\it Physica D} {\bf 196} 90-105  \vspace{-1.5mm}
\item Matsuno Y 2009  Soliton and periodic solutions of the short pulse model equation {\it Handbook of Solitons: Research, 
Technology and Applications} ed  S P Lang and  H Bedore (New York: Nova)  pp. 541-585  \vspace{-1.5mm}
\item Hone A N W,  Novikov V  and  Wang J P 2018 Generalizations of the short pulse equation {\it Lett. Math. Phys.} {\bf 108} 927-947   \vspace{-1.5mm}
\item Vakhnenko V A 1992 Solitons in a nonlinear model media {\it J. Phys. A: Math. Gen.} {\bf 25} 4181-4187   \vspace{-1.5mm}
\item Hunter J K  and  Saxton R 1991 Dynamics of director fields {\it SIAM J. Appl. Math.} {\bf 51} 1498-1521  \vspace{-1.5mm}
\item Sakovich S 2016 Transformation  and integrability of a generalized short pulse equation  {\it Commun. Nonlinear Sci. Numer. Simul.} {\bf 39} 21-28   \vspace{-1.5mm}
\item Matsuno Y 2016 Integrable multi-component generalization of a modified short pulse equation {\it J. Math. Phys.} {\bf 57} 111507   \vspace{-1.5mm}
\item Matsuno Y 2006 Cusp and loop soliton solutions of short-wave models of the Camassa-Holm and Degasperis-Procesi equations {\it Phys. Lett. A} {\bf 359} 451-457  \vspace{-1.5mm}
\item Manna M A   and  Neveu A 2001 Short-wave dynamics in the Euler equations {\it Inverse Problems} {\bf 17} 855-861  \vspace{-1.5mm}
\item Sakovich S Yu 2018 On a new avatar of the sine-Gordon equation {\it  Nonlinear Phenomena in Complex Systems} {\bf 21} 62-68   \vspace{-1.5mm}
\item Matsuno Y 2007 Multiloop soliton and multibreather solutions of the short pulse model equation  {\it J. Phys. Soc. Jpn.} {\bf 76} 084003  \vspace{-1.5mm}
\item Matsuno Y 2013 Smooth multisoliton solutions and their peakon limit of Novikov's Cammasa-Holm type equation  with cubic nonlinearity {\it J. Phys. A: Math. Theor.} {\bf 46} 365203 \vspace{-1.5mm}
\item Sawada K  and Kotera T 1974 A method for finding $N$-soliton solutions of the K.d.V. and K.d.V.-like equation {\it Prog. Theor. Phys.} {\bf 51} 1355-1367   \vspace{-1.5mm}
\item Dodd R K  and Bullough R K  1977  Polynomial conserved densities for the sine-Gordon equations {\it Proc. R. Soc. Lond. A} {\bf 352} 481-503   \vspace{-1.5mm}
\item Kraenkel R A, Leblond H  and Manna M A 2014 An integrable evolution equation for surface waves in deep water {\it J. Phys. A: Math. Theor.} {\bf 47} 025208  \vspace{-1.5mm}
\item Hirota R  and  Satsuma J 1976 $N$-soliton solutions of model equations for shallow water waves {\it J. Phys. Soc. Japan} {\bf 40} 611-612  \vspace{-1.5mm}
\item Li N  and  Qi H 2018 Multi-cuspon solutions of a generalized short pulse equation Preprint   \vspace{-1.5mm}
\item Feng B-F., Maruno K and Ohta Y 2012 On the $\tau$-functions of the reduced Ostrovsky equation and the $A^{(2)}_2$ two-dimensional Toda system {\it J. Phys. A: Math. Theor.} {\bf 45} 355203  \vspace{-1.5mm}
\item Matsuno Y 2010 A direct method for solving the generalized sine-Gordon equation II {\it J. Phys. A: Math. Theor.} {\bf 43} 375201  \vspace{-1.5mm}
\item Gerstner F 1809  Theorie der Wallen {\it Annl. Phys}. {\bf 32} 412-445   \vspace{-1.5mm}
\item Henry D 2008 On Gerstner's water wave  {\it J. Nonl. Math. Phys.} {\bf 15} Suppl 2  87-95  \vspace{-1.5mm}
\item Sakovich A and Sakovich S 2006 Solitary wave solutions of the short pulse equation {\it J. Phys. A: Math. Gen.} {\bf 39} L361-L367  \vspace{-1.5mm}
\item Belokolos E D, Bobenko A I, Enol'skii V Z, Its A R and Matveev V B 1994 {\it Algebro-
Geometric Approach to Nonlinear Integrable Equations} (Berlin: Springer)  \vspace{-1.5mm}
\item Gesztesy F and Holden H 2003 {\it Soliton Equations and Their Algebro-Geometric Solutions  {\rm vol} 1: (1+1)-Dimensional Continuous Models}({\it  Cambridge Stud. Adv. Math. {\bf 79}}) 
(Cambridge: Cambridge University Press)  \vspace{-1.5mm}
\item Guo Y and Yin Z 2019  Local well-posedness and blow-up phenomena of the generalized short pulse equation {\it  J. Math. Phys.} {\bf 60} 041505  \vspace{-1.5mm}
\item Matsuno Y 2011 A  novel multi-component generalization of the short pulse equation and its multisoliton solutions  {\it J. Math. Phys.}  {\bf 52} 123702   \vspace{-1.5mm}
\item Kraenkel R A, Leon J  and Manna M A 2005 Theory of small aspect ratio waves in deep water {\it Physica D} {\bf 211} 377-390  \vspace{-1.5mm}

\end{enumerate}
\end{document}